\documentclass[12pt]{iopart}
 \usepackage{graphicx} 

\usepackage{iopams}
\begin{document}

\title[Fluxes of atmospheric muons underwater]
{Fluxes of atmospheric muons underwater depending on the small-$x$
gluon density}


\author{A Misaki${}^1$, T S Sinegovskaya${}^2$, S I Sinegovsky${}^2$ \\ and N Takahashi${}^3$}

\address{${}^1$ Waseda University, Okubo 3-4-1, Shinjuku-ku, Tokyo,
169-8555 Japan}
\address{${}^2$ Irkutsk State University, 664003 Russia}
\address{${}^3$ Hirosaki University,  036-8561 Japan}

\ead{\mailto{sinegovsky@api.isu.runnet.ru}}

\begin{abstract}
The prompt muon contribution to the deep-sea atmospheric muon flux can serve as a
tool for probing into the small-x feature of the gluon density inside of a nucleon,
if the muon energy threshold could be lifted to 100 TeV. The prompt muon flux
underwater is calculated taking into consideration predictions of recent charm
production models in which the small-$x$ behaviour of the gluon distribution is
probed. We discuss the possibility of distinguishing the PQCD models of the charm
production differing in the small-x exponent of the gluon distribution, in
measurements of the muon flux at energies 10–100 TeV with neutrino telescopes.
\end{abstract}

\submitto{\JPG}


\section{Introduction}
A correct treatment of the charm hadroproduction is important to the
atmospheric muon and neutrino studies,  since  short-lived charmed
particles, $D^\pm$, $D^0$, $\overline{D}{}^0$, $D_s^\pm$,
$\Lambda_c^+$, which are produced in collisions of cosmic rays with
nuclei of the air, become the dominant source of atmospheric muons
and neutrinos at energies $E \sim 100$ TeV. Thus, one needs to take
them into consideration as the background for extraterrestrial
neutrinos (for a review, see \cite{mann}). Muons originating from
decay of these charmed hadrons are so called prompt muons (PM) that
contribute to the total atmospheric muon flux.

Another aspect of the interest to the charm production relates to the gluon density
at small gluon momentum fraction $x$. The gluon density at small $x$ is of
considerable importance because this strongly influences the charm production cross
section, both total and inclusive.  Recently Pasquali et al. \cite{prs99} and
Gelmini et al. \cite{ggv2,ggv3} have analysed the influence of small-$x$ behaviour
of the parton distribution functions (PDFs) on the atmospheric lepton fluxes at sea
level. Based on next-to-leading order (NLO) calculations of the perturbative
Quantum Chromodynamics (PQCD), they predict PM fluxes at the ground level depending
strongly on proton gluon distributions at small $x$ scale,  $x < 10^{-5}$.

The muon spectra underwater computed with the model of Pasquali et al.~\cite{prs99},
in which used were the MRSD$_-$ \cite{MRSD-} and the CTEQ3M \cite{CTEQ3} sets of
PDFs,  were recently  discussed \cite{27icrc,NSS00,note00}. In this note, using
predictions of the PQCD model \cite{ggv2,ggv3} for the charm production,  we discuss
the PM contribution to the deep-sea muon flux at depths typical for operating and
constructing neutrino telescopes, AMANDA~\cite{Amanda}, ANTARES~\cite{Antares},
Baikal~\cite{nt36}, NESTOR~\cite{Nestor}. Due to large detector volume and efective
area ($10^4-10^5$ m$^2$) and homogeneity of surrounding matter these underice and
deep-sea installations have considerable advantages over underground detectors for
probing very high-energy atmospheric muons.

Namely, here we try to study a PM flux underwater dependence on the power $\lambda$
of the small-$x$ gluon distribution function: $xg(x,Q^2)\propto x^{-\lambda}$. The
nature of the small-$x$ behaviour of the gluon density is now under extensive
discussion (see, for example, \cite{ander02,ball00,kaid01,schl,vogt00,yosh01}). The
small-$x$ behaviour of the PDFs is the subject of the deep interest because an
understanding of the underlying dynamics is far yet from being clear.

\section{PDFs and charm production models}
Due to dominant subprocess in heavy quarks hadroproduction,
$gg\rightarrow c\overline c$, the charm production is sensitive to
the gluon density at small $x$, where $x$ is the gluon momentum fraction.
One may evaluate the scale of $x$ in cosmic ray
interactions as follows.  The product of the gluon momentum fraction $x_{1}$
of the projectile nucleon and that of the target $x$ near
the charm production threshold ($\sim 2m_{c}$) is \,
\mbox{$x_{1}x=4m_{c}^{2}/(2m_{N}E_{0})$},\,\
 where $E_{0}$ is the primary
nucleon energy in the lab frame. Since a muon takes away about 5\% of
the primary nucleon energy, $E_{0}\simeq 20E_{\mu}$, we have
\mbox{$x_{1}x=0.1(m_{c}/m_{N})(m_{c}/E_{\mu})$.} Because of the
steepness of the primary cosmic ray spectrum only large $x_{1}$
contribute sizeably to the atmospheric charm production, so one needs
to adopt $x_{1}\gtrsim 0.1$. Taking $m_{c}^{2}\simeq 2\,$ GeV$^{2}$,
one may find the range of importance for $E_{\mu}\gtrsim 100$\,TeV to
be $x\lesssim 2\cdot10^{-6}$. It should be stressed, this range is
yet outside of the scope of the perturbative next-to-leading order
global analysis of parton distributions \cite{MRST,CTEQ5}.

The exponent $\lambda$ in PQCD charm production
models~\cite{prs99,ggv2,ggv3} covers wide range from about $0.5$, the
value being formerly connected to the Pomeron intercept $\Delta$ in
the leading order of the Balitsky-Fadin-Kuraev-Lipatov (BFKL)
approach~\cite{BFKL}, to about $0.1-0.2$ ($\Delta=0.13-0.18$), values
obtained with the NLO corrections~\cite{nloBFKL} to the BFKL scheme.
The interactions between Pomerons lead to the increase of the BFKL Pomeron
 intercept~\cite{kaid01}.

 \begin{figure}[t]
 \centering{
  \vskip -10mm
 \vspace*{2.0mm} 
 \includegraphics[width=10.0cm]{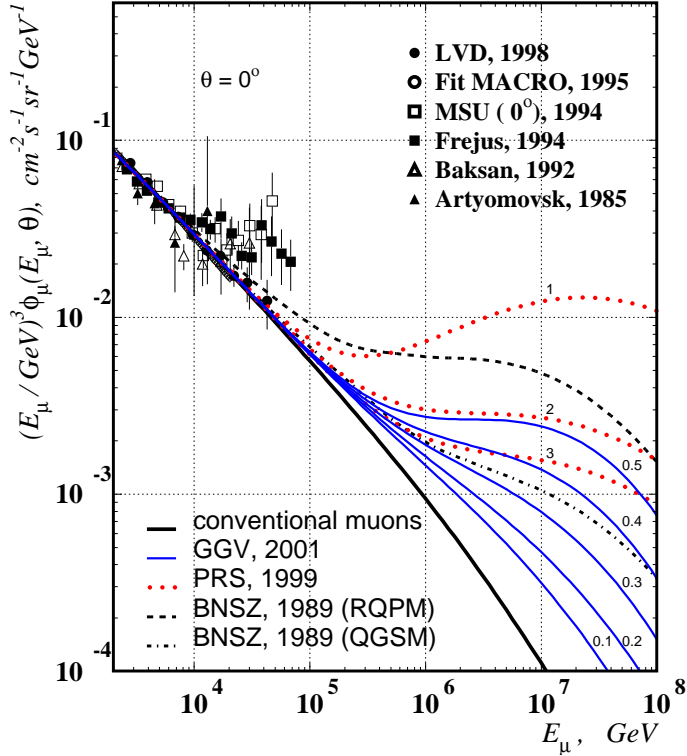}} 
\caption{Vertical sea-level muon flux data and predictions.
Experiments: $\blacktriangle$ -- Artyomovsk~\cite{ASD85}, $\triangle$
-- Baksan~\cite{BNO92}, $\opensquare$ -- MSU~\cite{MSU94},
$\fullsquare$ -- Frejus~\cite{Frejus94}, $\opencircle$ --
MACRO~\cite{MACRO95}, $\fullcircle$ -- LVD~\cite{LVD98}. The lower
solid line stands for conventional muons. The rest of curves
represent the total muon flux, sum of prompt muons and conventional
ones.
 \label{fig-1}}
  \vskip -3mm
 \end{figure}


Figure \ref{fig-1} presents the sea-level muon flux measured near the
vertical~\cite{ASD85,BNO92,MSU94,Frejus94,MACRO95,LVD98}, as well as fluxes
calculated with taking into account the prompt muon contribution. These ones are
predictions of the quark-gluon string model (QGSM)~\cite{olga,BNSZ89} (the
dash-dotted line) and the recombination quark-parton model
(RQPM)~\cite{BNSZ89,prd58} (dashed), as well those of a set of PQCD charm production
models by Pasquali, Reno and Sarcevic \cite{prs99} (hereafter PRS -- dotted lines
with numbers 1, 2, 3) and  models by Gelmini, Gondolo and Varieschi \cite{ggv2,ggv3}
(GGV -- thin curves with numbers 0.1 -- 0.5). These models are used further in
calculations of the deep-sea muon flux. Let us sketch out PQCD models.

\subsection{The model by Pasquali, Reno and Sarcevic}
\subsubsection{PRS-1.}

The PRS-1 model (dotted lines in figures~\ref{fig-1}, \ref{fig-2})
(identical with the PQCD-1 in reference~\cite{note00}) is based on
the MRSD$_-$ set \cite{MRSD-}. The PDF input parameters are the
followings: $xg(x,Q_0^2)\sim x^{-0.5}$ as $x\rightarrow0$, 4-momentum
transfer squared $Q_0^2=4$\,GeV$^2$; the sea light quark asymmetry,
$\overline u < \overline d$, is taking into consideration; the QCD
scale in the minimal subtraction scheme ($\overline {\rm MS}$),
$\Lambda^{\overline {\rm MS}}_4=0.215$\,GeV, corresponds to the
effective coupling at the $Z$ boson mass scale
$\alpha_{s}(M^{2}_{Z})=0.111$.  The factorization scale is
$\mu_F=2m_c$, the renormalization one is $\mu_R=m_c$, where the charm
quark mass, $m_c$, is chosen to be equal $1.3$. The sea-level prompt
muon flux has been parameterized by authors \cite{prs99} with the equation:
\begin{equation}\label{prs1}
\fl\lg[E^3_{\mu}\phi^{D,\,\Lambda_{c}}_{\mu}(E_{\mu})\cdot
(\mathrm{cm^{-2}s^{-1}sr^{-1}GeV^{2}})^{-1}]
=-5.91+0.290y+0.143y^{2}-0.0147y^{3},
 \end{equation}
\noindent
 where $y=\lg(\frac{E_{\mu}}{1 \, \rm GeV}).$

\subsubsection{PRS-2.}

 In the PRS-2 model (the same as the PQCD-2 in reference \cite{note00})
 CTEQ3M set~\cite{CTEQ3} was used.
Corresponding inputs which were utilized in this model are
$\Lambda^{\overline{MS}}_4=0.239$\, GeV, 
$\alpha_{s}(M^{2}_{Z})=0.112,$ \,
 $m_c=$ 1.3 GeV, \,
$\mu_F=2m_c$,  $\mu_R=m_c$,\, and  $\lambda=0.286$\, at
$Q^{2}_{0}=1.6\,{\rm GeV}.$
 The corresponding approximate expression for the PM spectrum is

\begin{equation}\label{prs2}
\fl \lg[E^{3}_{\mu}\phi^{D,\,\Lambda_{c}}_{\mu}(E_{\mu})\cdot
(\mathrm{cm^{-2}s^{-1}sr^{-1}GeV^{2}})^{-1}]=-5.79+0.345y+0.105y^{2}-0.0127y^{3}.
\end{equation}

\subsubsection{PRS-3.}

In this model the CTEQ3M set was also used. Differing  from PRS-2 in the
renormalization and factorization scales, $\mu_F=\mu_R=m_c$, this model shows the
uncertainty relating to the scale choice. In this case the PM spectrum was given as

\begin{equation}\label{prs3}
\fl\lg[E^3_{\mu}\phi^{D,\,\Lambda_{c}}_{\mu}(E_{\mu})\cdot
(\mathrm{cm^{-2}s^{-1}sr^{-1}GeV^{2}})^{-1}]=-5.37+0.0191y+0.156y^{2}-0.0153y^{3}.
\end{equation}

\subsection{The model by Gelmini, Gondolo and Varieschi}

 Here we present results for the model, among those
discussed in \cite{ggv3}, which is based on MRST set of  PDFs
\cite{MRST} with different values of the exponent $\lambda$ in the
range $0.1-0.5$, \,$Q^{2}\geq 1.25$\,GeV$^{2}$;\,
$\alpha_{s}(M^{2}_{Z})=0.1175.$ The factorization and renormalization
scales are:
$$
  \mu_F=2m_T, \,\mu_R=m_T, \,
$$
where
$$
m_{T}=(k_{T}^{2}+m_c^{2})^{1/2},\,
  m_c= 1.25\, {\rm GeV},
$$
and characteristic transverse momentum $k_{T}$
is of $\sim m_c.$

In order to compute PM flux underwater we parameterize  sea-level muon
spectra of the GGV model (see figure 7 in reference~\cite{ggv3}) with
the formulae:

\begin{equation}\label{pms}
\phi_{\mu}^{D,\,\Lambda_{c}}(E_\mu)
 = A\left(\frac{E_\mu}{1\,\rm GeV}\right)^{-(\gamma_{0}+\gamma_{1}y+
\gamma_{2}y^{2}+\gamma_{3}y^{3})}
\mathrm{cm^{-2}s^{-1}sr^{-1}GeV^{-1}}.
\end{equation}
In table~\ref{t1} five sets of the parameters to equation~(\ref{pms})
are presented for different values of the index $\lambda$ of the
small-$x$ gluon destribution.
\begin{table}[h!]
\caption{Parameters of the prompt muon
 spectrum at sea level (\ref{pms}).
\label{t1}}

\begin{indented}
\lineup
\item[]\begin{tabular}{llllll}
 \br
 $\lambda$ &\m$A, 10^{-6}$&\m$\gamma_{0}$&\m$\gamma_{1}$&\m$\gamma_{2},
 10^{-2}$&\m$\gamma_{3}, 10^{-3}$\cr
 \mr
 $0.1$&\m$3.12$&\m$2.70$&\m$\-0.095$&\m$ 1.49$&\m$\-0.2148$ \cr
 $0.2$&\m$3.54$&\m$2.71$&\m$\-0.082$&\m$ 1.12$&\m$\-0.0285$ \cr
 $0.3$&\m$1.80$&\m$2.38$&\m$0.045$&\m$\-0.82$&\m$0.911$ \cr
 $0.4$&\m$0.97$&\m$2.09$&\m$0.160$&\m$\-2.57$&\m$1.749$ \cr
 $0.5$&\m$0.58$&\m$1.84$&\m$0.257$&\m$\-4.05$&\m$2.455$ \cr
 \br
\end{tabular}
\end{indented}
\end{table}

\section{The conventional muon flux}

The main source of the atmospheric muons up to $\sim 50 $ TeV are
decays of secondary cosmic ray pions and kaons. The flux
(conventional) of ($\pi, K$)-muons  is computed based on the nuclear
cascade model by \cite{VNS86} (see also \cite {prd58,NSS98}).
High-energy part of this spectrum for the vertical may be
approximated with the equation (in
$\mathrm{cm^{-2}s^{-1}sr^{-1}GeV^{-1}}$):

\begin{equation}\label{our}
 \phi_\mu^{\pi,K}\left(E_\mu,0^\circ \right) =
  \cases{14.35\,E_\mu^{-3.672} &for $E_1<E_\mu\leqslant E_2$\,, \\
 10^{3}E_\mu^{-4} &for $E_\mu>E_2$\\}.
 \end{equation}
where $E_1=1.5878\times10^3\,\mathrm{GeV}, E_2=4.1625\times10^5$\,GeV.

Zenith-angle distribution of atmospheric muons at sea-level was
computed in the reference~\cite{Tanya99} where detail comparison
between the calculated atmospheric muon spectra and the sea-level
experimental data at different zenith angles was made (see
also~\cite{note00}). The conventional muon flux computed for the
vertical direction is shown in figure~\ref{fig-1} (the lower solid
line).

\begin{figure}[t!]
 \centering{
 \vskip -8mm
 \vspace*{2.0mm} 
 \includegraphics[width=10.0cm]{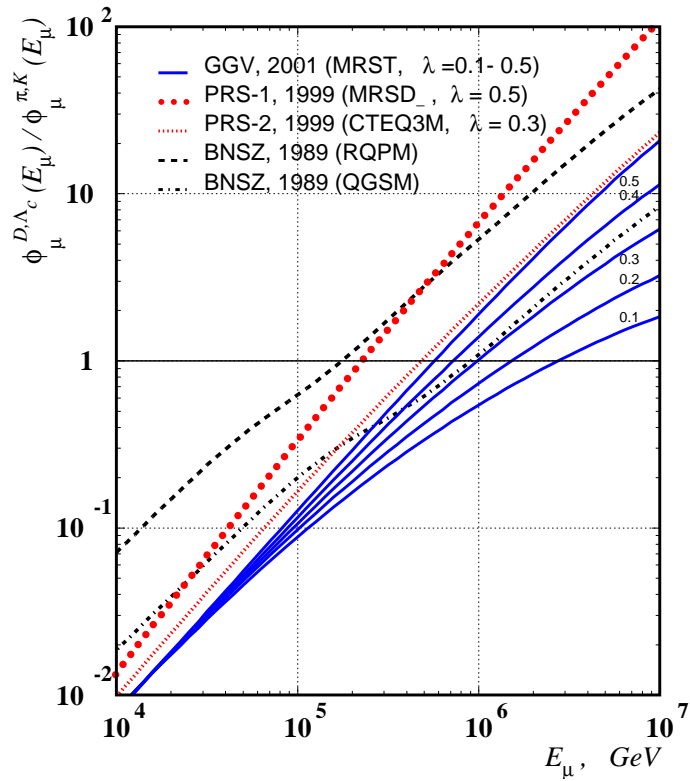}} 
\vskip -2mm
 \caption{Ratio of the differential prompt muon spectrum at sea level
 to the  conventional one.
 \label{fig-2}}
 \end{figure}

Each of five thin lines in figure~\ref{fig-1} presents the sum of the
conventional muon flux~(\ref{our}) and the GGV prompt muon
flux~(\ref{pms}) corresponding to the exponent $\lambda=0.1,\, 0.2,\,
0.3,\, 0.4,\, 0.5$ (numbers near lines). Dotted lines show the same
for PRS models, equations~(\ref{prs1}-\ref{prs3}).
For comparison there are also shown contributions due to the quark-gluon string
model and the recombination quark-parton one \cite{BNSZ89,prd58} (the dash-dot line
and the dash line respectively). Ratios of prompt muon fluxes to the conventional
one are shown in figure~\ref{fig-2}. As one can see, the crossover energy for the PM
flux and conventional one covers the wide region from $\sim ~ 150$\,TeV to $\sim ~ 3
$\,PeV, that is more than one order of the magnitude.

It is worth to note that old QGSM prediction~\cite{BNSZ89} at
 high energies is within GGV prompt muon fluxes  as well that of RQPM is
within PRS results (figures~\ref{fig-1}, \ref{fig-2}).

\section{Prompt muon component of the flux  underwater}

Muon energy spectra and angle distributions of the flux underwater
was computed with the method by~\cite{NSB94}. The collision integral
in the kinetic equation includes  the energy loss of muons due to
brems\-strah\-lung, direct $e^+e^-$ pair production and photonuclear
interactions. The ionization energy loss and the small-$v$ part of
the loss due to $e^+e^-$ pair production ($v < 2\cdot10^{-4}$, where
$v$ is the fraction of the energy lost by the  muon)  were treated as
continuous ones.

In our calculations of underwater muon fluxes at different zenith
angles, we used, as a boundary spectra, PQCD PM fluxes calculated
only for the vertical direction at the ground level, supposing the
isotropic approximation for prompt muons to be a reliable at least
for $10^4< E_\mu < 10^6\,$ GeV at zenith angles $\theta\lesssim
80^\circ.$

\begin{figure} [t!]
 \centering{
 \vskip -17mm
 \vspace*{2.0mm} 
 \includegraphics[width=9.0cm]{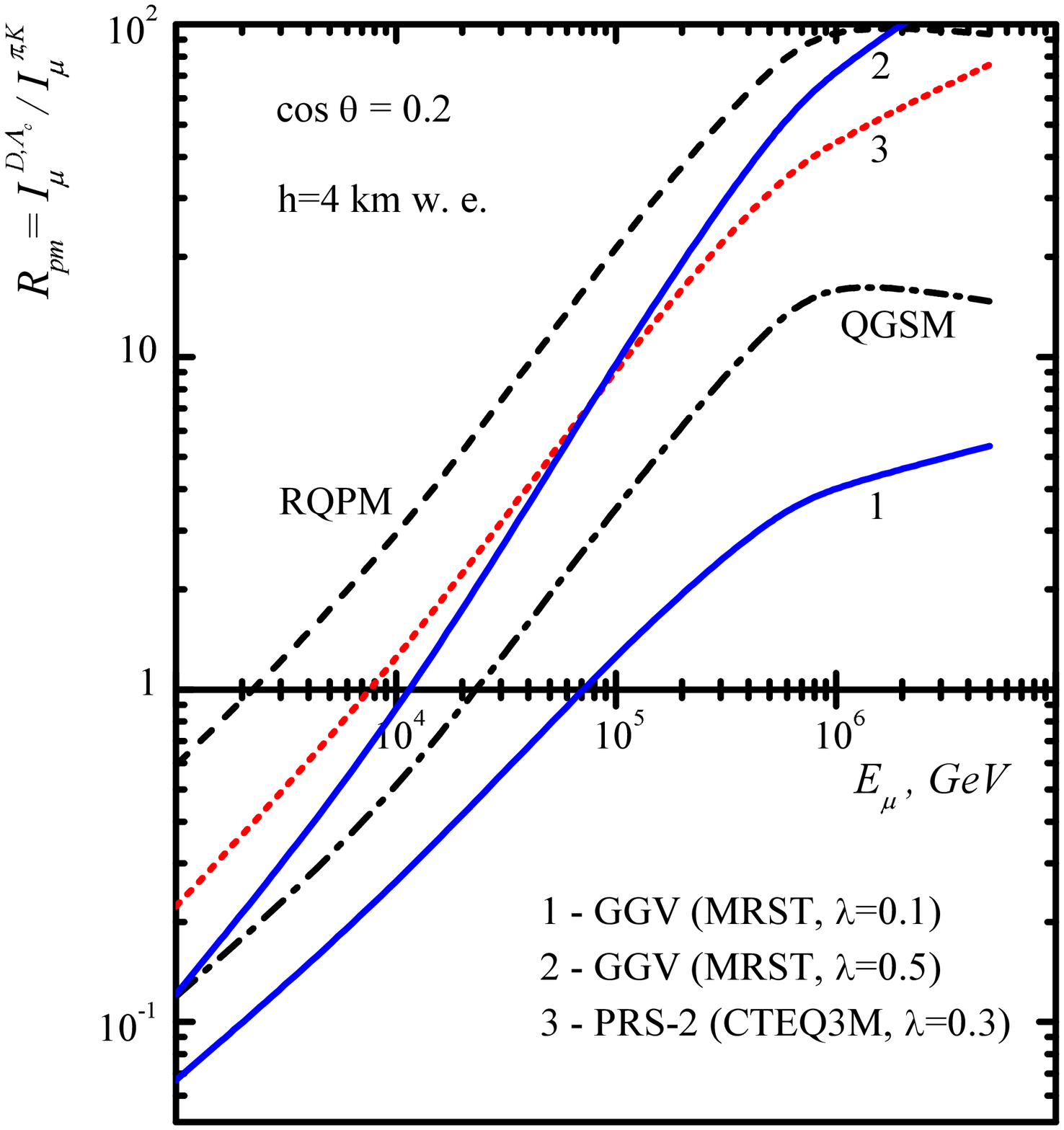}} 
 \vskip -19mm
 \caption{Prompt muon contribution at $h=4$\,km w. e. vs. $E_{\mu}$.
   \label{fig-3}}
 \vskip -0 mm
    \end{figure}
 \begin{figure}
 \centering{
  \vskip -15mm
 \vspace*{2.0mm} 
  \includegraphics[width=9.0cm]{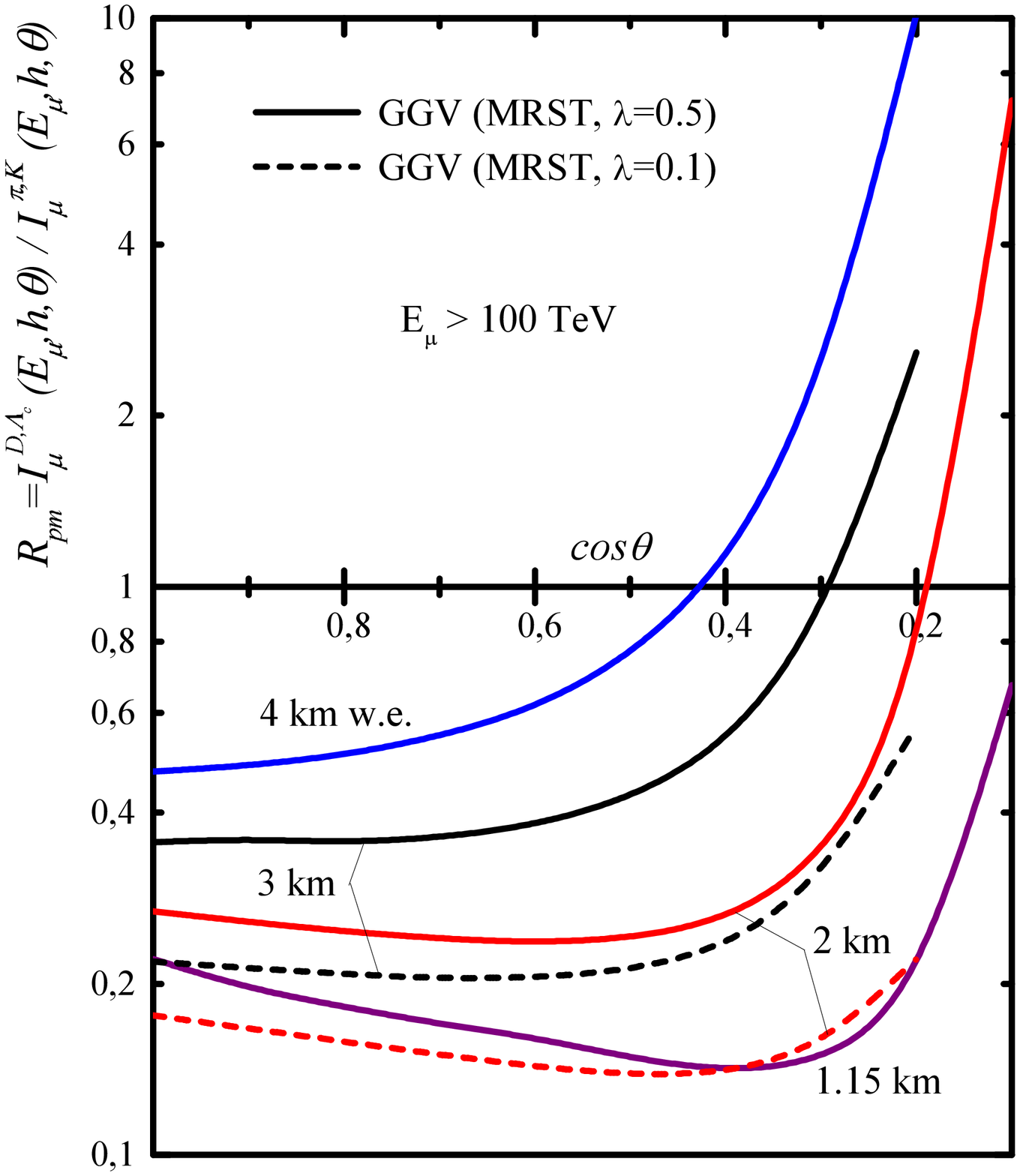}} 
 \vskip -15mm
 \caption{Ratio of the prompt muon flux underwater
   to the conventional one as a function of $\cos\theta$
   at $E_{\mu}\ge 100$\,TeV.
   \label{fig-4}}
   \end{figure}

The prompt muon fraction of the flux underwater, $R_{pm}$, defined as ratio of the
prompt muon integral spectrum to the conventional one, is presented in
figure~\ref{fig-3} for the depth of 4 km of the water equivalent (w. e.) and for
$\cos\theta=0.2$. As is seen from this figure, $R_{pm}$ related to the gluon density
slope $\lambda=0.5$ is a factor 3 greater than that for $\lambda=0.1$ at $E_\mu
\gtrsim 10$ TeV.

Zenith-angle distributions of the prompt muon contribution at depths $1$-$4$\,km
w.~e., calculated for $E_\mu > 100$\,TeV,  are shown in figure~\ref{fig-4}. Here we
used predictions of the GGV model for two values of the gluon density exponent,
 $\lambda=0.1$ (dash) and $\lambda=0.5$ (solid).
As one can see in figure~\ref{fig-4}, $R_{pm}$ increases for the vertical direction
from about 0.2 at the depth of the Baikal NT (1.15 km)~\cite{nt36} to about 0.5 at
the NESTOR depth ($\sim 4$ km)~\cite{Nestor}.  For the larger zenith angles,
$\theta\sim 75^{\circ}$, this contribution becomes apparently sizable at depths
$3-4$\,km. Differences in the predictions owing to a change of $\lambda$, from $0.1$
to $0.5$ (see $h=2$ and $3$ km w. e.), are also clearly visible: the ratio
$R_{pm}(\lambda=0.5)/R_{pm}(\lambda=0.1)$ at $h=2$\,km w. e. grows from about $1.5$
to about $5$ as $\cos\theta$ changes from $1$ to $0.2$.

Here we supposed no differences between PRS and GGV calculations
apart from those related to the charm production cross sections.
Actually one needs  to compare the primary spectrum and composition,
nucleon and meson production cross sections and other details of the
atmospheric nuclear cascade being used in above computations. These
sources of uncertainties would be considered elsewhere.

\section{Summary}
In order to test the small-$x$ gluon distribution effect we have computed deep-sea
prompt muon fluxes using predictions of charm production models based on NLO
calculations of the PQCD \cite{prs99}-\cite{ggv3}.
The possibility to discriminate the PQCD models, differing in the
slope of the gluon distribution, seems to be achievable in
measurements of the underwater muon flux at energies 50-100 TeV.

Hardly appeared at sea level for energies up to $10^5$ GeV (figures~\ref{fig-1},
\ref{fig-2}), a dependence on the spectral index $\lambda$ of the small-$x$ gluon
distribution becomes more distinct at depths $3-4$\,km w. e. (figures~\ref{fig-3},
\ref{fig-4}). At the depth of $4$\, km and at the angle of $\sim 78^{\circ}$ one
could observe the PM flux to be equal, for $\lambda=0.5$, to the conventional one
even for muon energy $\sim 10$\,TeV (the crossover energy). While for $\lambda=0.1$
the crossover energy is about $70$\,TeV. For the high energy threshold,
$E_{\mu}>100$\,TeV, and at $h\lesssim 3$\, km w. e., the ratio $R_{pm}$ is nearly
isotropic up to $\sim 60^{\circ}$. The ``crossover zenith angle'' at a given depth,
$\theta_c(h)$, depends apparently on the small-$x$ exponent $\lambda$ of the gluon
density inside colliding nucleons:
$$\cos\theta_{c}\mid_{\lambda=0.5}\,\simeq0.3 \ \ \mathrm{and} \
 \ \cos\theta_{c}\mid_{\lambda=0.1}
\,\simeq 0.1 \ \ \mathrm{for} \ h=3\,\, \mathrm{km\, w.\, e.}$$


\section*{References}



\begin{thebibliography}{99}

\bibitem{mann}
Learned J  G and Mannheim K 2000 {\it Ann. Rev. Nucl. Part. Sci.} {\bf 50}, 679.

\bibitem{prs99}
Pasquali L, Reno M H and Sarcevic I 1999 {\it Phys. Rev.} D {\bf 59}, 034020.

\bibitem{ggv2}
Gelmini G, Gondolo P and  Varieschi G 2000
 {\it Phys. Rev.} D {\bf 61}, 056011.

\bibitem{ggv3}
Gelmini G,  Gondolo P and Varieschi G 2001 {\it Phys. Rev.} D {\bf 63}, 036006.

\bibitem{MRSD-}
Martin A D, Stirling  W J and  Roberts R G 1993 {\it Phys. Rev.} D {\bf 47}, 867.

\bibitem{CTEQ3}
 Lai H L et al. 1995 {\it Phys. Rev.} D {\bf 51}, 4763;
 Lai H L et al. 1997 {\it Phys. Rev.} D {\bf 55}, 1280.

\bibitem{27icrc}
 Misaki A. et al 1999 {\it Proc.\ 26 ICRC (Salt Lake City)}
        vol~2, p~139, hep-ph/9905399.

\bibitem{NSS00}
 Naumov  V A, Sinegovskaya T S and Sinegovsky S I 2000
{\it Phys. Atom. Nucl.} {\bf 63}, 1923.

\bibitem{note00}
 Sinegovskaya T S and Sinegovsky S I 2001
 {\it Phys. Rev.}  D {\bf 63}, 096004.

\bibitem{Amanda}
  Andres E et al. (AMANDA Collaboration) 2000
{\it Astropart. Phys.} {\bf 13}, 1.

\bibitem{Antares}
 Amram P et al. (ANTARES Collaboration) 2000
{\it Astropart. Phys.} {\bf 13}, 127.

\bibitem{nt36}
Belolaptikov I  A et al. (Baikal Collaboration) 1997 {\it Astropart. Phys.} {\bf 7},
263.

\bibitem{Nestor}
Anassontzis  E G et al. (NESTOR Collaboration) 2000 {\it Nucl. Phys. B (Proc.
Suppl.)} {\bf 85}, 153.

\bibitem{ander02}
Andersson B (Small x Collaboration) 2002 Small x phenomenology:
 Summary and status, \\ hep-ph/0204115.

\bibitem{ball00}
 Ball R D and Landshoff P V 2000 \jpg {\bf 26}, 672.

\bibitem{kaid01}
Kaidalov A B 2001 Regge poles in QCD,  hep-ph/0103011.

\bibitem{schl}
Schleper P 2001 Soft hadronic interactions, hep-ex/0102051.

\bibitem{vogt00}
 Vogt R 2000 {\it Prog. Part. Nucl. Phys.} {\bf 45}, S105.

\bibitem{yosh01}
Yoshida R (on behalf of ZEUS and H1 Collaboration) 2001 HERA small-$x$ and/or
diffraction,  hep-ph/0102262.

\bibitem{MRST}
Martin A D, Roberts R G, Stirling  W J and Torne R S 1999 {\it Nucl. Phys. B (Proc.
Suppl.)} {\bf 79}, 105.

\bibitem{CTEQ5}
Lai H L et al 2000 Eur. Phys. J. C {\bf 12}, 375.

\bibitem{BFKL}
 Kuraev E A, Lipatov L N and Fadin V S 1976 {\it Zh. Eksp. Teor. Fiz.} {\bf 71}, 840;
\item[] Kuraev E A, Lipatov L N and Fadin V S 1977 {\it Zh. Eksp. Teor. Fiz.} {\bf 72}, 377;
\item[] Balitsky I I and Lipatov L N 1978 {\it Yad. Fiz.} {\it Yad. Fiz.} {\bf 28}, 1597.

\bibitem{nloBFKL}
 Brodsky S J et al 1999 {\it JETP Lett.} {\bf 70}, 155,  hep-ph/9901229;
 \item[] Kim V T, Lipatov L N and Pivovarov G B 1999 The Next-to-Leading dynamics of the
 BFKL Pomeron, hep-ph/9911242.

\bibitem{ASD85}
        Khalchukov F F et al 1985 {\it Proc.\ 19 ICRC (La Jolla)}
        vol~8, p~12.

\bibitem{BNO92}
 Bakatanov V N et al. 1992 {\it Yad. Fiz.} {\bf 55}, 2107.

\bibitem{MSU94}
  Zatsepin  G T et al. Bull. of the Russian Acad. of  Sci. Ser. Phys.
  1994 {\bf 58}, 2050.

\bibitem{Frejus94}
 Rhode W 1994 {\it Nucl. Phys.}  B (Proc. Suppl.) {\bf 35}, 250.

\bibitem{MACRO95}
Ambrosio M et al (MACRO Collaboration) 1995
    {\it Phys. Rev.} D {\bf 52}, 3793.

\bibitem{LVD98}
Aglietta M et al (LVD Collaboration) 1998
 {\it Phys. Rev.} D {\bf 58}, 092005;
\item[]Aglietta M et al (LVD Collaboration)1999
 {\it Phys.Rev.} D {\bf 60}, 112001.

\bibitem{olga}
Kaidalov A B 1982 {\it Phys. Lett.} B {\bf 116}, 459;
\item[] Kaidalov A B and Piskunova O I 1986 {\it Sov. J. Nucl. Phys.} {\bf 43}, 1545.

\bibitem{BNSZ89}
 Bugaev E V et al 1989 \NC C {\bf 12}, 41.

\bibitem{prd58}
 Bugaev E V et al 1998 \PR D {\bf 58}, 054001.

\bibitem{VNS86}
 Vall  A N, Naumov  V A and Sinegovsky S I 1986
{\it Sov. J. Nucl. Phys.} 1986 {\bf 44}, 806.

\bibitem{NSS98}
Naumov V A, Sinegovskaya T S and Sinegovsky S I 1998 \NC A {\bf 111}, 129.

\bibitem{Tanya99}
         Sinegovskaya T S 1999 {\it Proc. Second Baikal
         School on Fundamental Physics ``Interaction of Radiation
         and Fields with Matter''} (Irkutsk: Irkutsk University Press)
         vol~2, p~598 (in Russian).

\bibitem{NSB94}
Naumov V A, Sinegovsky S I and Bugaev E V 1994 {\it Phys. Atom. Nucl.} {\bf 57},
412; hep-ph/9301263.

\end{thebibliography}
\end{document}